\documentclass[a4paper,12pt]{article}
\usepackage{amsmath}
\usepackage{amsfonts}
\usepackage{amssymb}
\usepackage{amsthm}
\usepackage{amscd}
\usepackage{array}
\allowdisplaybreaks[3]

\title{The Asymptotic Behavior of Bouncing Cosmological Models in $F(\mathcal{G})$ Gravity Theory}

\author{
Andrey N. Makarenko$^{1}$\footnote{e-mail:\ andre@tspu.edu.ru}
and
Alexander N. Myagky$^{2}$\footnote{e-mail:\ alexmyagky@gmail.com}\\
{\it\small $^1$Tomsk State Pedagogical University, ul. Kievskaya, 60, 634061, Tomsk, Russia}\\
{\it\small $^2$National Research Tomsk Polytechnic University, Lenin Avenue, 30, 634050, Tomsk, Russia}}

\date{}

\begin{document}

\maketitle

\begin{abstract}
We reconstruct $F(\mathcal{G})$ gravity theory with an exponential scale factor
to realize the bouncing behavior in the early universe and
examine the asymptotic behaviour of late-time solutions in this model.
We propose an approach to the construction of asymptotic expansions of solutions
of the Friedmann equations on the basis of Puiseux series.
\end{abstract}



\section{Introduction}

The Big Bang era is one of the less understood periods of the evolution of our Universe,
and the physics behind this era is still inconceivable. The classical cosmological approach
leads inevitably to an initial singularity, which is a rather "embarrassing" feature
of the classical description, because due to this singularity, the closed time-like geodesics
which pass from this singularity, have a finite proper length, but no end points to normal
space away from the singularity. However, not so long ago there was an alternative
description - the matter bounce scenario~\cite{Brand:2010a,Brand:2010b,Brand:2011,Brand:2012}.
In this scenario, in the contraction phase the universe
is dominated by matter, and a non-singular bounce occurs. Also, the density perturbations
whose spectrum is consistent with the observations
can be produced (for a review, see~\cite{Novello:2008ra}).
In addition, after the contracting phase,
the so-called BKL instability~\cite{Belinsky:1970ew}
happens, so that the universe will be anisotropic.
The way of avoiding this instability~\cite{Erickson:2003zm} and
issues of the bounce~\cite{Xue:2010, Xue:2011} in the Ekpyrotic scenario~\cite{Khoury:2001wf}
has been investigated~\cite{Cai:2012va, Cai:2013vm, Qiu:2013}.
Moreover, the density perturbations
in the matter bounce scenario with two scalar fields
has recently been examined~\cite{Cai:2013kja}.
On the other hand, various cosmological observations support
the current cosmic accelerated expansion.
To explain this phenomenon in the homogeneous and isotropic universe,
it is necessary to assume the existence of dark energy, which
has negative pressure, or propose that gravity is modified on large scales
(for recent reviews on issues of dark energy and modified gravity theories,
see, e.g.,~\cite{Nojiri:2010wj, Nojiri:2006ri, Bamba:2014eea, Capozziello:2010, Capozziello:2011et, Bamba:2012cp}).
Regarding the latter approach, there have been proposed a number of
modified gravity theories such as $F(R)$ gravity.
Bounces in modified gravity of F(R) type mainly have been studied
in ~\cite{Nojiri:2016ygo, Odintsov:2015ynk, Odintsov:2015zza, Haro:2015mmo, Odintsov:2015zua, Odintsov:2015uca}.
A relation between the bouncing behavior and
the anomalies on the cosmic microwave background radiation also been discussed~\cite{Liu:2013kea}.

The asymptotic behavior of bouncing models is interesting due to
various reasons. First of all, it may give the information about
$\Lambda$CDM era of bounce cosmology~\cite{Cai:2014, Cai:2015,
Odintsov:2015zua, Cai:2016, DBounce}. Second, it maybe important
for the understanding of possible presence of weak singularities
at the very-early and very-late bounce universe. Indeed, it is
known that bounce universe maybe still weakly singular, see
examples in~\cite{Odintsov:2015ynk, Odintsov:2015zza}. The study
of weak singularities in bouncing universe may suggest the way to
remove finally all the singularities and construct the totally
regular bounce. Finally, the asymptotic behavior of solutions may
indicate new, not yet explored possibilities for the universe
evolution.


\section{$F(\mathcal{G})$ theory of gravity}

In this paper, we explore bounce cosmology in $F(\mathcal{G})$ gravity.
This class of modified gravity is based on the use of the Gauss-Bonnet
invariant
$\mathcal{G}=R^{2}-4R_{\mu\nu}R^{\mu\nu}+R_{\mu\nu\rho\sigma}R^{\mu\nu\rho\sigma}$,
where $R_{\mu\nu}$ is the Ricci tensor and $R_{\mu\nu\rho\sigma}$ is
the Riemann tensor.

The action of $F(\mathcal{G})$ gravity model is described as \cite{Nojiri:2005jg}
\begin{equation}
S=\frac{1}{2\kappa^2}\int d^4 x\sqrt{-g}\left(R+F(\mathcal{G})\right)+S_{\mathrm{matter}}\,,
\label{eq:1}
\end{equation}
where $g$ is the determinant of the metric tensor $g_{\mu\nu}$
and $S_{\mathrm{matter}}$ is the matter action.
We use units of $k_\mathrm{B}=c_{\mathrm{l}}=\hbar=1$,
where $c$ is the speed of light, and denote the
gravitational constant $8\pi G$ by ${\kappa}^2\equiv 8\pi/M_{\mathrm{Pl}}^2$
with the Planck mass of $M_{\mathrm{Pl}}=G^{-1/2}=1.2\times 10^{19}$\,\,GeV.

It follows from this action that the gravitational field equation reads
\begin{align*}
R_{\mu\nu}
&-\frac{1}{2}g_{\mu\nu}R-\frac{1}{2}g_{\mu\nu}F(\mathcal{G})+\\
&+\left(
2RR_{\mu\nu}-4R_{\mu\rho}R_{\nu}{}^{\rho}
+2R_{\mu}{}^{\rho\sigma\tau}R_{\nu\rho\sigma\tau}
-4g^{\alpha\rho}g^{\beta\sigma}R_{\mu\alpha\nu\beta}R_{\rho\sigma}
\right)F'(\mathcal{G})-\\
&-2\left({\nabla}_{\mu}{\nabla}_{\nu}F'(\mathcal{G})\right)R
+2g_{\mu\nu}\left(\Box F'(\mathcal{G})\right)R
-4\left(\Box F'(\mathcal{G})\right)R_{\mu\nu}+\\
&+4\left({\nabla}_{\rho}{\nabla}_{\mu}F'(\mathcal{G})\right)R_{\nu}{}^{\rho}
+4\left({\nabla}_{\rho}{\nabla}_{\nu}F'(\mathcal{G})\right)R_{\mu}{}^{\rho}
-4g_{\mu \nu}\left({\nabla}_{\rho}{\nabla}_{\sigma}F'(\mathcal{G})\right)R^{\rho\sigma}+\\
&+4\left({\nabla}_{\rho}{\nabla}_{\sigma}F'(\mathcal{G})\right)
g^{\alpha\rho}g^{\beta\sigma}R_{\mu\alpha\nu\beta}
=\kappa^2T^{(\mathrm{matter})}_{\mu\nu}.
\label{eq:2}
\end{align*}

Here, the prime denotes the derivative with respect to
$\mathcal{G}$, ${\nabla}_{\mu}$ is the covariant derivative, $\Box
\equiv g^{\mu \nu}{\nabla}_{\mu}{\nabla}_{\nu}$ is the covariant
d'Alembertian, and
\begin{equation*}
T^{\mu(\mathrm{matter})}_{\,\,\,\,\nu}=
\mathrm{diag}\left(-\rho_{\mathrm{matter}},p_{\mathrm{matter}},p_{\mathrm{matter}},p_{\mathrm{matter}}\right)
\end{equation*}
is the energy-momentum tensor of matter, where
$\rho_{\mathrm{matter}}$ and $p_{\mathrm{matter}}$ are the energy density
and pressure of matter, respectively.

We take the flat Friedmann-Lema\^{i}tre-Robertson-Walker (FLRW) metric,
given by
\begin{equation*}
ds^2=-dt^2+a^2(t)\sum_{i=1,2,3}\left(dx^i\right)^2\,,
\label{eq:3}
\end{equation*}
where $a$ is the scale factor,
$H=\dot{a}/a$ is the Hubble parameter, and
the dot shows the time derivative.
In this background, we have $R=6\dot{H}+12H^2$
and $\mathcal{G}=24H^2\left(\dot{H}+H^2\right)$.
The gravitational field equations become~\cite{Cognola:2007vq}
\begin{equation}
6H^2+F(\mathcal{G})-\mathcal{G}F'(\mathcal{G})
+24H^3\dot{\mathcal{G}}F''(\mathcal{G})=2\kappa^2\rho_{\mathrm{matter}},
\label{eq:4}
\end{equation}
\begin{align*}
4\dot{H}
&+6H^2+F(\mathcal{G})-\mathcal{G}F'(\mathcal{G})
+16H\dot{\mathcal{G}}\left(\dot{H}+H^2\right)F''(\mathcal{G})+\\
&+8H^2\ddot{\mathcal{G}}F''(\mathcal{G})+8H^2\dot{\mathcal{G}}^2F'''(\mathcal{G})
=-2\kappa^2p_{\mathrm{matter}}.
\label{eq:5}
\end{align*}
In what follows, we investigate only gravity part of the action
in Eq.~(\ref{eq:1}) without its matter part.

We examine the following form of the scale factor
\begin{equation}
a(t)=\exp\left(\alpha t^2\right),\quad \alpha>0.
\label{eq:6}
\end{equation}
Here $\alpha$ is a constant with the dimension of mass squared
$([Mass]^2)$. From this expression we have
\begin{equation}
H(t)=2\alpha t,\quad
\mathcal{G}(t)=192t^2\alpha^3(1+2t^2\alpha).
\label{eq:7}
\end{equation}
We should note that $\mathcal{G}\geq 0$ for any $t$.

From Eq.~(\ref{eq:7}) we see that a cosmological bounce happens in
the early universe at the time $t=0$. On the other hand, when
$\alpha t^2\gg 1$, the universe can be considered to be at the
dark energy dominated stage, because taking into account
Eq.~(\ref{eq:6}), we get
$$\Ddot{a}(t)=2\alpha(1+2\alpha t^2)\exp(\alpha t^2)>0.$$
This implies the accelerated expansion of the universe happens.
Thus we see that in the case when the scale factor is given by Eq.~(\ref{eq:6})
the late-time cosmic acceleration as well as the bouncing behavior
in the early universe can be realized in a unified manner.



\section{Reconstruction method of $F(\mathcal{G})$ gravity}

Next, we reconstruct $F(\mathcal{G})$ gravity models
by using the method \cite{Capozziello:2006dj, Nojiri:2006gh, Bamba:2008ut}.
Introducing proper functions $P(t)$ and $Q(t)$ of a scalar field $t$,
which is interpreted as the cosmic time,
the action in Eq.~({\ref{eq:1}}) without matter is described as
\begin{equation}
S=\frac{1}{2\kappa^2}\int d^{4}x\sqrt{-g}\left(R+P(t)\mathcal{G}+Q(t)\right)\,.
\label{eq:8}
\end{equation}
By varying this action with respect to $t$, we obtain
\begin{equation}
\frac{dP(t)}{dt}\mathcal{G}+\frac{dQ(t)}{dt}=0.
\label{eq:81}
\end{equation}
Solving this equation in terms of $t$, we get $t=t(\mathcal{G})$.
The substitution of $t=t(\mathcal{G})$ into Eq.~(\ref{eq:8}) yields
$F(\mathcal{G})=P(t)\mathcal{G}+Q(t)$.
Using this equation and Eq.~(\ref{eq:4}), we find
\begin{equation}
Q(t)=-6H^2(t)-24H^3(t)\frac{dP(t)}{dt}.
\label{eq:9}
\end{equation}
With this equation and the relation $F(\mathcal{G})=P(t)\mathcal{G}+Q(t)$,
we acquire
\begin{equation}
2H^2\frac{d^2P(t)}{dt^2}+2H\left(2\dot{H}-H^2\right)
\frac{dP(t)}{dt}+\dot{H}=0.
\label{eq:10}
\end{equation}

Suppose the scale factor is given by Eq.~(\ref{eq:6}),
the general solution of Eq.~(\ref{eq:10}) becomes
\begin{align*}
P(t)=
&c_1
+c_2\left(2\alpha\,t\,{}_1F_1\left(\frac{1}{2};\frac{3}{2};\alpha t^2\right)-\frac{1}{t}e^{\alpha t^2}\right)-\\
&-\frac{1}{12}t^2\,{}_2F_2\left(1,1;2,\frac{5}{2};\alpha t^2\right)
-\frac{1}{8\alpha}\ln(\alpha t^2),
\end{align*}
where $c_1$ and $c_2$ are arbitrary constants,
${}_1F_1\left(a;b;z\right)$ and ${}_2F_2\left(a_1,a_2;b_1,b_2;z\right)$ are generalized hypergeometric functions.
From Eq.~(\ref{eq:9}), we obtain
\begin{equation*}
Q(t)=-24\alpha^2t^2-192c_2\alpha^{3}te^{\alpha t^2}
+48\alpha^2t^2e^{\alpha t^2}\,{}_1F_1\left(\frac{1}{2};\frac{3}{2};-\alpha t^2\right).
\end{equation*}
Plugging this expression with Eq.~(\ref{eq:81}), we have
\begin{equation}
t=\pm\frac{1}{4\sqrt{3}\alpha}\sqrt{-12\alpha+\sqrt{6}\sqrt{G+24\alpha^2}},\quad \mathcal{G}\geq 0.
\label{eq:14}
\end{equation}
Accordingly, by solving $F(\mathcal{G})=P(t)\mathcal{G}+Q(t)$, we find
the most general form of $F(\mathcal{G})$ as
\begin{multline}
F(\mathcal{G})=c_1\mathcal{G}
+c_2\left(2\alpha\,t\,\mathcal{G}\,{}_1F_1\left(\frac{1}{2};\frac{3}{2};\alpha t^2\right)
-(\mathcal{G}+192t^2\alpha^3)\frac{1}{t}e^{\alpha t^2}\right)
-24\alpha^2t^2+\\
+48\alpha^2t^2e^{\alpha t^2}{}_1F_1\left(\frac{1}{2};\frac{3}{2};-\alpha t^2\right)
-\frac{1}{12}t^2\mathcal{G}\,{}_2F_2\left(1,1;2,\frac{5}{2};\alpha t^2\right)
-\frac{1}{8\alpha}\mathcal{G}\ln(\alpha t^2),
\label{eq:15}
\end{multline}
where $t$ determined by the expression (\ref{eq:14}).


\section{Asymptotic behaviour of solutions}

We explore the exponential form of the scale factor (\ref{eq:6}) and
for this case Friedmann equation (\ref{eq:4}) has the form
\begin{equation}
p_2(\mathcal{G})\frac{d^2 F(\mathcal{G})}{d\mathcal{G}^2}
+p_1(\mathcal{G})\frac{d F(\mathcal{G})}{d\mathcal{G}}
+F(\mathcal{G})=b(\mathcal{G}),
\label{eq:33}
\end{equation}
with
\begin{align*}
p_2(\mathcal{G})=
&-192\alpha^2\mathcal{G}-16\alpha(\mathcal{G}+48\alpha^2)(12\alpha-\sqrt{6}\sqrt{\mathcal{G}+24\alpha^2}),\\
p_1(\mathcal{G})=
&-\mathcal{G},\\
b(\mathcal{G})=
&\frac{1}{2}\left(12\alpha-\sqrt{6}\sqrt{\mathcal{G}+24\alpha^2}\right).
\end{align*}
It is not difficult to see that the differential equation has two
singularities. One of them ($\mathcal{G}=0$) is a regular
singularity and another ($\mathcal{G}=\infty$) is an irregular
singularity.

One can consider the homogeneous equation cor\-res\-pon\-ding to Eq.~(\ref{eq:33})
\begin{equation}
\frac{d^2F(\mathcal{G})}{d\mathcal{G}^2}
+q_1(\mathcal{G})\frac{dF(\mathcal{G})}{d\mathcal{G}}
+q_2(\mathcal{G})F(\mathcal{G})=0.
\label{eq:37}
\end{equation}
Here the coefficients $q_1(\mathcal{G})$ and $q_2(\mathcal{G})$ have the following form
\begin{equation*}
q_1(\mathcal{G})=\frac{p_1(\mathcal{G})}{p_2(\mathcal{G})},\quad q_2(\mathcal{G})=\frac{1}{p_2(\mathcal{G})}.
\end{equation*}
It is obviously that $F(\mathcal{G})=\mathcal{G}$ is a solution of Eq.~(\ref{eq:37}).

We seek a solution of Eq.~(\ref{eq:33}) in the neigh\-bor\-hood of $\mathcal{G}=0$.
First of all, we construct a fundamental system of solutions of the homogeneous equation (\ref{eq:37}).
Since the coefficients $q_k (\mathcal{G})$ for $k=1, 2$ has a pole of order not higher
than $k$ at $ G=0$, then we can obtain
\begin{equation*}
q_1(\mathcal{G})=\frac{\bar{q}_1(\mathcal{G})}{\mathcal{G}},
\quad
q_2(\mathcal{G})=\frac{\bar{q}_2(\mathcal{G})}{\mathcal{G}^2},
\label{eq:38}
\end{equation*}
where $\bar{q}_1(\mathcal{G})$ and $\bar{q}_2(\mathcal{G})$ are
holomorphic functions in a neighborhood of $\mathcal{G}=0$.
We construct a fundamental system of solutions of Eq.~(\ref{eq:37})
in the neighborhood of $\mathcal{G}=0$.
Solutions will be found in the form of a generalized series
\begin{equation*}
F(\mathcal{G})=\mathcal{G}^\mu\sum_{k=0}^{\infty}A_k\mathcal{G}^k.
\label{eq:39}
\end{equation*}
By combining this expression with Eq.~(\ref{eq:37}), we acquire
\begin{equation}
\mu\left(\mu-1\right)+\bar{q}_1(0)\mu+\bar{q}_2(0)=0.
\label{eq:40}
\end{equation}
By solving this equation, we get $\mu_1=1/2$ and $\mu_2=1$.
The value $\mu=1$ corresponds to solution $F(\mathcal{G})=\mathcal{G}$.
The second solution of Eq.~(\ref{eq:37}) we represent
\begin{equation*}
F(\mathcal{G})=\sqrt{\mathcal{G}}\varphi(\mathcal{G}),
\label{eq:41}
\end{equation*}
where $\varphi(\mathcal{G})$ is holomorphic function in the neigh\-bor\-hood of $\mathcal{G}=0$
at that $\varphi(\mathcal{G})\neq 0$. Substituting Eq.~(\ref{eq:40}) into the homogeneous equation,
we obtain a recurrent system from which we consistently find the coefficients $A_0$, $A_1$, $\ldots$.
Thus, the fundamental system of the homogeneous equation has the form
\begin{equation*}
F_1(\mathcal{G})=\mathcal{G},
\end{equation*}
\begin{equation*}
F_2(\mathcal{G})=
\sqrt{\mathcal{G}}\left(1-\frac{\mathcal{G}^2}{2^{12}3^3\alpha^4}
+O(\mathcal{G}^3)\right),\quad
\mathcal{G}\to 0.
\end{equation*}
By solving the inhomogeneous equation (\ref{eq:33}) using the method of variation of constants,
we obtain an approximate solution at $\mathcal{G}\to 0$
\begin{equation*}
F(\mathcal{G})=
c_1\mathcal{G}
+c_2\sqrt{\mathcal{G}}
\left(1-\frac{\mathcal{G}^2}{2^{12}3^3\alpha^4}
+O(\mathcal{G}^3)\right)
-\frac{1}{8\alpha}\mathcal{G}\ln\mathcal{G}
+\frac{\mathcal{G}^2}{2^83^2\alpha^3}
+O(\mathcal{G}^3),
\label{eq:45}
\end{equation*}
where $c_1$ and $c_2$ are arbitrary constants.
In addition, it should be mentioned that we could have obtained the same result
if we expand Eq.~(\ref{eq:15}) in a fractional power of $\mathcal{G}$.

Next, we will construct an asymptotic expansion of solution of Eq.~(\ref{eq:33}) at $\mathcal{G}\to\infty$.
As above, we find a fundamental system of solutions of homogeneous equation (\ref{eq:37}),
where the coefficients of equation are represented as asymptotic series
\begin{equation*}
q_1(\mathcal{G})=
-\frac{1}{16\sqrt{6}\alpha}\frac{1}{\mathcal{G}^{1/2}}-\frac{1}{4\mathcal{G}}
-\frac{3\sqrt{6}\alpha}{8}\frac{1}{\mathcal{G}^{3/2}}+\ldots,\quad
q_2(\mathcal{G})=-\frac{1}{\mathcal{G}}\,q_1(\mathcal{G}).
\label{eq:21}
\end{equation*}
Asymptotic solutions will be found in the form of a Puiseux series~\cite{Olver:1974}
\begin{equation*}
F(\mathcal{G})=
\exp(\lambda\mathcal{G}^{1/2})\mathcal{G}^{\sigma}\sum_{k=0}^{\infty}A_k\mathcal{G}^{-k/2}
\label{eq:399}
\end{equation*}
By combining this expression with Eq.~(\ref{eq:37}), we acquire
\begin{equation*}
\lambda\left(\lambda-\frac{1}{8\sqrt{6}\alpha}\right)=0,
\quad
\sigma=\frac{\sqrt{6}-36\alpha\lambda}{\sqrt{6}-96\alpha\lambda}.
\end{equation*}
Solutions of this equations are $\lambda_1=0$, $\sigma_1=1$
and $\lambda_2=1/(8\sqrt{6}\alpha)$, $\sigma=-1/4$.
The first of them is correspond to solution $F(\mathcal{G})=\mathcal{G}$.
Therefore the asymptotic expansion of the second solution of Eq.~(\ref{eq:37})
can be represented in the form
\begin{equation}
F(\mathcal{G})=\exp\left(\frac{\mathcal{G}^{1/2}}{8\sqrt{6}\alpha}\right)\varphi(\mathcal{G}),
\label{eq:46}
\end{equation}
where $\varphi(\mathcal{G})$ is Puiseux asymptotic series at $\mathcal{G}\to\infty$.
Substituting Eq.~(\ref{eq:46}) into the homogeneous equation,
we obtain a recurrent system from which we consistently find
the coefficients $A_0$, $A_1$, $\ldots$.
Thus, the asymptotic expansion for solution of Eq.~(\ref{eq:33})
has the form
\begin{equation*}
F(\mathcal{G})=
\exp\!\left(\frac{\mathcal{G}^{1/2}}{8\sqrt{6}\alpha}\right)\!\mathcal{G}^{-1/4}
\!\left(1
+O\!\left(\frac{1}{\mathcal{G}^{1/2}}\right)\!\right),\ \ \mathcal{G}\to\infty.
\label{eq:48}
\end{equation*}
Asymptotic expansion for solution of inhomogeneous equation (\ref{eq:33}) will be found in the form
\begin{equation}
F(\mathcal{G})=
\mathcal{G}\sum_{k=0}^{\infty}A_k\mathcal{G}^{-k/2}.
\label{eq:49}
\end{equation}
Substituting Eq.~(\ref{eq:49}) into Eq.~(\ref{eq:33}) we find the coefficients $A_k$.
As a result, we obtain
\begin{equation*}
F(\mathcal{G})=
A_0\mathcal{G}-\sqrt{6}\mathcal{G}^{1/2}-18\alpha+O\left(\frac{1}{\mathcal{G}^{1/2}}\right),\quad
\mathcal{G}\to\infty,
\label{eq:50}
\end{equation*}
where $A_0$ is arbitrary constant.

Note that there is another asymptotic expansion for the solution of Eq.~(\ref{eq:33})
\begin{align*}
F(\mathcal{G})=
&A_0\mathcal{G}-\sqrt{6}\mathcal{G}^{1/2}-18\alpha+O\!\left(\frac{1}{\mathcal{G}^{1/2}}\right)+\\
&+B_0\exp\left(\frac{\mathcal{G}^{1/2}}{8\sqrt{6}\alpha}\right)\mathcal{G}^{-1/4}
\!\left(1
+O\!\left(\frac{1}{\mathcal{G}^{1/2}}\right)\!\right),
\quad\!\!\mathcal{G}\to\infty,
\label{eq:51}
\end{align*}
where $A_0$ and $B_0$ are arbitrary coefficients.
Varying these coefficients leads to the asymptotic behaviour change.
In particular, it is easy to see that if we take $A_0=0$ and $B_0=0$, the asymptotic behavior
solutions of Eq.~(\ref{eq:33}) will be determined by $\mathcal{G}^{1/2}$.
We can prove that this fact corresponds to the choice of a specific value of the constant
$c_1=0$ and $c_2=-\sqrt{\pi}/(8\alpha^{3/2})$ in exact solution (\ref{eq:15}).

In addition, consider bouncing cosmological models with scale factor
\begin{equation*}
a(t)=\exp(\alpha t^2)+\exp(\alpha^2t^4),\quad \alpha>0.
\end{equation*}
It is not easy to reconstruct function $F(\mathcal{G})$ for this model.
However is quite possible, to investigate the asymptotic behavior of solutions of Friedmann equation.
In this case, the differential equation (\ref{eq:33}) has two singularities: $\mathcal{G}=0$
(regular point) and $\mathcal{G}=\infty$ (irregular point).
The solution in the neighborhood of $\mathcal{G}=0$ has been investigated in the work \cite{Bamba:2014my}.
On the basis of Puiseux series we construct an asymptotic expansion of the solution when $\mathcal{G}\to\infty$
\begin{align*}
F(\mathcal{G})=
&A_0\mathcal{G}-\sqrt{6}\mathcal{G}^{1/2}
-\frac{2^{7/6}3^{23/6}}{5}\alpha^{2/3}\mathcal{G}^{1/6}
+O\!\left(\frac{1}{\mathcal{G}^{1/6}}\!\right)\!+\\
&+B_0\exp\left(\frac{\mathcal{G}^{1/3}}{2^{11/3}3^{1/3}\alpha^{2/3}}\right)\mathcal{G}^{-1/12}
\!\left(1
+O\!\left(\frac{1}{\mathcal{G}^{1/3}}\right)\!\right).
\end{align*}

\section{Summary}

We have reconstructed $F(\mathcal{G})$ gravity model with ex\-po\-nen\-ti\-al scale factor
and found that in this model the bouncing behavior can happen.
Also, we have explored the behavior of solutions of Friedmann equations
for this model at the singularities of the differential equation.
In particular, the Puiseux series were used to obtain the asymptotic expansion at an irregular singularity.
In addition, it has been verified that in a sum of two exponential functions model of the scale factor,
asymptotic behavior at late-time cosmic acceleration is similar to that of the model discussed above.
Recently there was an paper \cite{DBounce} one can demonstrated that in the context of LQC,
it is possible to realize a deformed matter bounce scenario,
in which the deformation practically alters the late-time behavior of the model.
Would be interesting to apply the proposed mechanism for the constructed in our paper models.

\section*{Acknowledgments}

This work was supported by a grant of the Russian Ministry of Education and
Science.


\end{document}